%% file: Satellite.tex
\documentclass[final]{IEEEtran}
\usepackage[update,prepend]{epstopdf}

\usepackage{graphics}
\usepackage{multirow}
\usepackage{tikz}
\usepackage{bbm} 
\usepackage{pdfpages}
\usepackage{multirow}
\usepackage{subfig}
\usepackage{comment}

\usepackage{setspace}	
\usepackage{graphicx}
\usepackage{algorithm,algorithmic}
\usepackage{multicol}

\usepackage[justification=centering]{caption}
\usepackage{textcomp}
\usepackage{psfrag}
\usepackage{arydshln}
\usepackage{url}
\usepackage{soul}
\usepackage{graphicx,color}
\usepackage[nolist]{acronym}

\usepackage{mathtools,lipsum}
\usepackage{cuted}
\usepackage{amsmath}
\usepackage{graphicx}

\include{notation}

\begin{document}
\graphicspath{{./Figures/}}
	\input{Elzanaty_Acronyms.tex}	
	\input{Definitions}

\title{
Ultra-Dense LEO Satellite-based Communication Systems: A Novel Modeling Technique
}

\author{Ruibo Wang, Mustafa A. Kishk, and Mohamed-Slim Alouini
\thanks{Ruibo Wang, Mustafa A. Kishk and Mohamed-Slim Alouini are with King Abdullah University of Science and Technology (KAUST), CEMSE division, Thuwal 23955-6900, Saudi Arabia} 
\vspace{-4mm}}

\maketitle
\begin{abstract}
Low earth orbit (LEO) satellite plays an indispensable role in the earth network because of its low latency, large capacity, and seamless global coverage. For such an unprecedented extensive irregular system, stochastic geometry (SG) is a suitable research method. The SG model can not only cope with the increasing network scale but also accurately analyze and estimate the network's performance. Several standard satellite distribution models and satellite-ground channel models are investigated in this paper. System-level metrics such as coverage probability and their intermediates are introduced in the non-technical description. Then, the influence of some factors on latency and coverage probability is studied. Finally, this paper presents the possible challenges and corresponding solutions for the SG-based LEO satellite system analysis.
\end{abstract}

\section{Introduction}
In the last few years, we have been witnessing a boom in the industry of LEO satellite networks. Constellations such as Starlink, Telesat, and OneWeb are planning to or have already deployed thousands of satellites \cite{del2019technical}. With the relatively good quality in terms of capacity and latency, LEO satellites provide an affordable solution for seamless network coverage. However, a strong and tractable mathematical framework is still needed to study and analyze the performance of LEO satellite-based communication systems. In particular, with a system composed of large numbers of LEO satellites deployed at different altitudes from the earth, analysis will enable us to understand the required numbers and altitudes of satellites to maximize system performance. In addition, it will enable us to to learn how much earth stations are needed in order to enable seamless operation of the LEO satellite communication system.
\par
Based on the above description, the structure of this paper is as follows. First, we describe the advantages and disadvantages of low-orbit satellites in more detail. Then we give the motivation for applying SG to LEO satellite networks. Based on SG, this paper goes over the analysis framework of some recent literature and gives a non-technical description of models of satellite locations distribution, channel model, typical parameters, and main performance metrics. In addition, a simulation setup is proposed to highlight the potential gains of using SG in modeling and analyzing LEO satellite communication systems with useful system-level insights concluded. Finally, we predict future scenarios for the challenges of applying SG in satellite networks.

\begin{figure*}[h]
	\centering
	\includegraphics[width=0.99\linewidth]{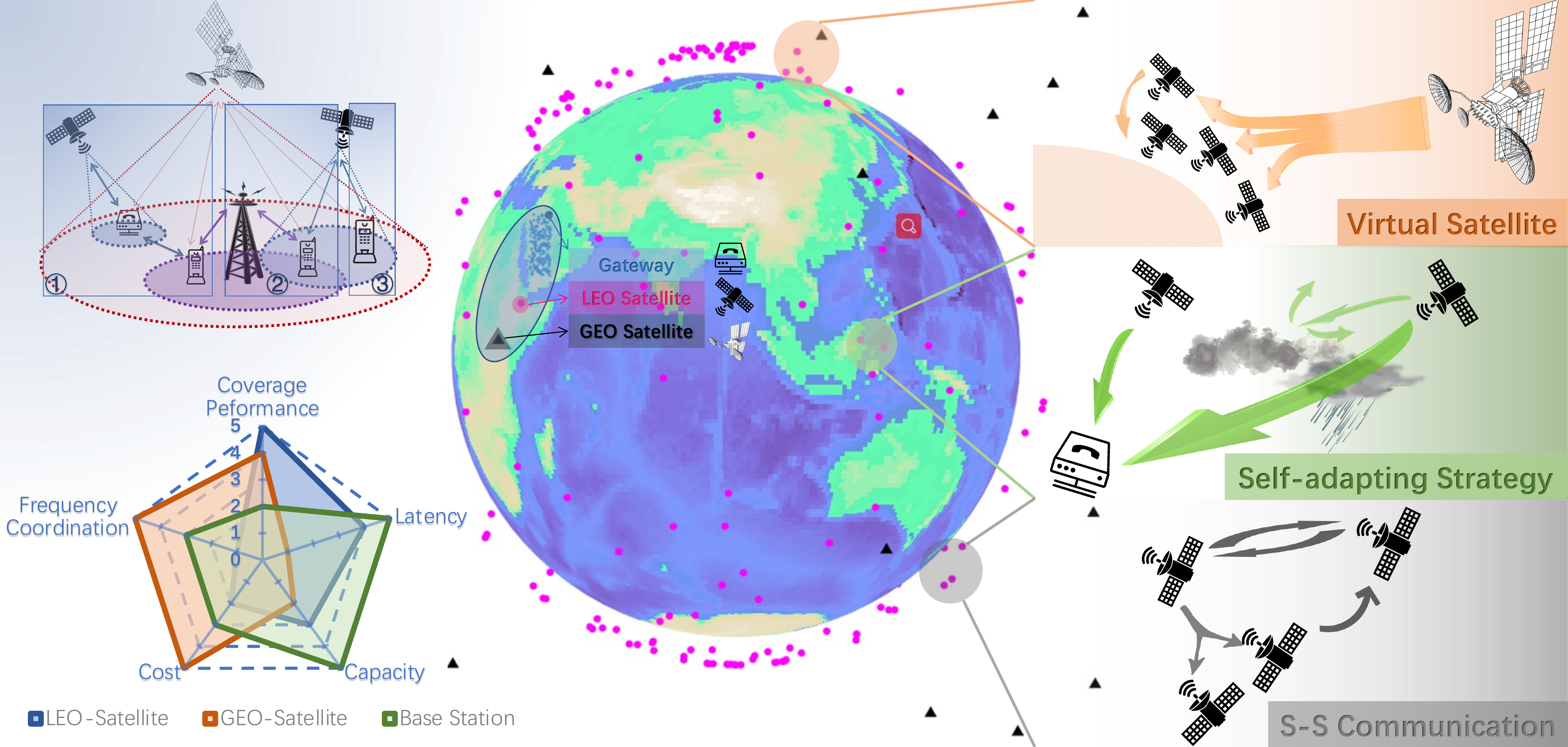}
	\caption{System-level metric comparison of three systems and three future scenarios.}
	\label{fig:roadmap}
\end{figure*}
\par

\subsection{Advantages and Challenges of LEO Satellite System}
In this section, the advantages and challenges of the LEO satellite system are analyzed by comparing with the geosynchronous equatorial orbit (GEO) satellite system and the terrestrial system \cite{liu2021leo}. The radar diagram in the lower-left corner of Fig.~\ref{fig:roadmap} visually shows the qualitative comparison points.
\par
LEO satellite system can provide seamless coverage of the earth and has incomparable advantages in providing reliable coverage. Despite the enormous coverage of a single GEO satellite, the system has blind areas at the poles. The coverage of terrestrial systems is limited by terrain. In remote areas that lack infrastructure, existing ground networks are challenged to provide coverage because the cost of extending fiber optics to these areas is expensive.
\par
LEO satellite system has relatively larger capacity density and lower latency. In the same region, LEO satellite systems provide much more capacity than GEO satellite systems, but less than terrestrial systems. The capacity is sufficient to support high rate communications in rural and remote areas, but not enough to meet the needs of densely populated cities. The latency of LEO satellites ranges from several milliseconds to tens of milliseconds, which can meet the needs of most application scenarios. Application scenarios requiring high real-time performance, such as the Internet of vehicles, are still highly dependent on BSs. The long communication distance of the GEO satellite system leads to a much longer delay than the others. 
\par
However, LEO satellite systems face significant challenges in terms of cost and frequency coordination. A large LEO satellite system requires high construction and maintenance costs. Fortunately, aerospace manufacturers have brought down launch costs, and streamlined production is bringing down the cost of producing satellites. Due to the low altitude of the terrestrial system, a large number of base stations need to be deployed to ensure coverage. While construction costs are higher, maintenance costs of the ground stations are lower than satellites because they are located on the ground. The smaller GEO satellite system certainly has a cost advantage. In terms of frequency licensing, the International Telecommunication Union (ITU) gives frequency priority to GEO satellites. Because of the first-claim, first-possession principle, once these huge constellations are built, it is not easy to allocate dedicated spectrum to the new constellations. There is also some channel overlap between LEO satellite system and 5G ground base station system, which further complicates the challenges of spectrum resource allocation problems. 

\subsection{Motivations of Applying SG in LEO Satellite Network}
As a powerful mathematical tool, SG has been widely used in cellular networks. It also has advantages in LEO satellite system modeling and analysis.  
\par
The existing satellite analysis methods are mainly to design the specific constellation. But for such a large system, fine-grained modeling of each LEO satellite would take a lot of effort. It is far more tractable to do SG-based system level analysis than to define and study the behavior of each satellite. SG modeling does not need to rely on orbital metrics and the shape of a specific constellation. A universal and flexible random network is what the explosive growth of dynamic networks requires. 
\par
For satellite systems, more satellites and lower altitudes mean larger density. In this case, interference will be an important factor affecting system performance. However, apart from SG, there is still a lack of practical analytical tools to simulate the interference caused by satellites. 
\par
In the existing non-SG models, satellites are often deployed in regular circular or hexagonal cells with the same coverage \cite{mourad2001generalized}. However, the distribution of satellites depends on latitude, and the coverage of satellites varies significantly in most cases. Therefore, the results of traditional analysis methods may greatly differ from the actual situation. SG is suitable for modeling and analyzing irregular topological networks \cite{haenggi2012stochastic}. In recent studies, various types of point processes used in LEO satellite network allow multiple methods to simulate the distribution. 
\par
Applying SG requires the assumption that satellites are independent and therefore can only be used to study the lower bounds of system performance. However, it has been proved that the lower bound of coverage probability and average achievable rate in the SG model is as tight as the upper bound of the regular mesh model \cite{andrews2011tractable}. For the LEO satellite system, the performance of the SG model closely matches with that of the deterministic constellation in coverage probability and average achievable rate \cite{OK_1}.

\section{SG-based Analytical Framework}
In this section, we go over the analytical framework  based on SG in recent works. Table~\ref{table:models} at the end of this section provides a brief summary of these frameworks. Before introducing the model, we first introduce some definitions in astronomy.

\begin{itemize}
    \item \textbf{Latitude}: Latitude is the line plane angle between the normal line on the earth and the equatorial plane. It ranges from 0 degrees at the equator to 90 degrees at the poles.
    \item \textbf{Zenith Angle}: Zenith angle is the angle between the incident signal and the direction of the local zenith (The geographical vertical).
    \item \textbf{Horizon}: The horizon is a two-dimensional plane tangent to the earth through the perception point, which can be the location of the gateway or the user. From the gateway's perspective, any satellite below the horizon is beyond line of sight since the earth acts as a blockage.
\end{itemize}

\subsection{Models of Satellite Locations Distribution}
Before studying satellite distribution, it is assumed that satellites, as well as users, satellite gateways (GW), and BSs are independent and uniformly distributed in the existing literature. The following sections will discuss the three-point process distribution of satellite networks.
\par
For \textit{Poisson Point Process (PPP)}, sample points obtained by random sampling obey uniform distribution, and the number of points in any given area is Poisson distributed while their locations are uniformly distributed within this area. As the most frequently used point process, PPP can fit the ground network well in performance analysis such as coverage probability and achievable data rate. Therefore, the positions of the LEO satellite are modeled as a PPP on a spherical surfaces of fixed height \cite{Al_1}. One of the most significant advantages of PPP is that it can predict the probability of a certain number of points in a spherical region with a specific area measure.
\par
Although PPP has a strong practicality, it is not suitable for modeling a finite area network with limited nodes. Since the positions of satellites are on a spherical surfaces, and for deterministic constellations, the number of satellites is fixed. Therefore, modeling the position of the satellite as a \textit{Binomial Point Process (BPP)}  \cite{OK_1}, \cite{MU_2} is an effective solution. Specifically, it places a fixed number of points on the fixed height sphere. The zenith (elevation) and azimuth angle are uniformly distributed.
\par
\textit{Non-Homogeneous Poisson Point Process (NPPP)}
improves the PPP model from another perspective. Since the distribution of deterministic satellite constellations is not uniform at different latitudes, that is, the number of satellites at the limit of inclination is actually greater than that in the equatorial region, the number of effective satellites per latitude in NPPP is used to compensate for the uneven density \cite{OK_3}. 
\par
Corresponding to the labels in the upper left section of Fig.~\ref{fig:roadmap}, there are three typical scenarios: GW-relayed system, terrestrial station and LEO satellite hybrid network and satellite to user direct communication scenario. The GW can provide directional transmission to the satellite. By deploying a GW as a relay, satellite coverage can be significantly improved \cite{MU_1}. Satellite networks can also assist ground networks. In addition to further increasing system capacity, it can also provide communication during emergencies. Satellites can also communicate directly with users, which of course requires special user equipment to enable direct communication. Since the coverage radius of GW or BS is negligible compared to the surface area of the earth, GWs and BSs can be regarded as points located in a large two-dimensional plane, and PPP is suitable to model their locations.

\subsection{Channel Model}
The linear expression of the received power expressed as $\rho \, \eta \, G \, r^{-\alpha}$, where $\rho$ is used to describe the transmitted power and antenna gain, $\eta$ and $G$ are for Large scale fading and Small scale fading, and $\alpha$ is called the path-loss exponent.
\par
There are two main models of signal power transmission. Take the downlink transmission as an example. The first is Equivalent Isotropically Radiated Power (EIRP) criteria \cite{Al_1}, that is, a satellite has the same gain in all directions. The second is a satellite that has a larger gain in a particular direction, in which the GW or user can receive main-lobe power, which can be realized by beam-forming technology \cite{Al_4}. Due to directional transmission, the received signal from the interfering source tends to come from the side-lobe, which can improve the \ac{SIR} of the system. In addition, in the current literature, satellites at the same altitude are typically assumed to have equal and constant transmission power.
\par
The path-loss exponent is a number between $-4$ and $-2$ usually. Consider that satellite-to-ground link propagates in free space in most cases, $\alpha=-2$ is satisfied for open areas. However, when the ground clutter tier reflects and absorbs the power, the multi-path effect cannot be ignored, resulting in $\alpha=-4$. An intuitive result is that when $\alpha=-2$, the interference becomes the main factor limiting system performance, and when $\alpha=-4$, the influence of noise will be far more than that of interference.
\par
Large scale fading is usually modeled as a log-normal shadowing to describe additional gain. It is inversely proportional to the square of carrier frequency \cite{OK_1}, \cite{Al_1}, \cite{MU_2}. Furthermore, some literature take the attenuation of air absorption caused by the resonance of water vapor into consideration \cite{Al_4}, \cite{MU_2}, which is called rain attenuation. It changes with the geographical location, such as desert and rain forest. 
\par
Small scale fading is a random variable which denotes the channel fading power gains. In ideal free space propagation, Small scale fading does not occur. Non-fading helps to investigate the upper bound on system performance. On the contrary, Rayleigh fading helps investigate the lower bound, which happens when serious multi-path distortion occurs. Rician fading is suitable for the general situation and is widely applied to make accurate performance estimates. Shadowed-Rician (SR) fading is a Rician fading channel with fluctuating LoS components. It is the most accurate channel fading model, especially for space-to-ground links \cite{MU_2}.
\par
Handling \ac{NLoS} interference is one of the keys to the accuracy of a channel model. Satellites below the horizon may not provide reliable service, but the diffraction signal may still interfere with the user. At the same time, the satellite may not establish a line-of-sight link due to obstacles from the ground. The former is often regarded as zero or some small constant rather than a random variable because the signal is relatively weak. The latter can be described by SR fading. It has been proved by simulation that SR fading can accurately simulate the channel model under \ac{NLoS} interference \cite{alfano2007sum}. In addition, multiplying by a factor representing the loss of redundant paths in Large scale fading is also an effective solution. This factor is modeled as a Log-Normal mixture random variable with two components, and the mixing ratio is given by the probabilities of \ac{LoS} and NLoS \cite{Al_1}. Notice that the probability that an LoS link is established is only determined by the zenith angle.

\subsection{Contact Distance and Satellite Availability}
Contact distance is the distance between the user/GW and the associated (service-providing) satellite.
Based on the strongest average received power association strategy, the contact distance is the distance from the user to the nearest satellite at a given altitude. As a random variable, the \ac{CCDF} of contact distance at $d_0$ is equal to the probability that there are no satellites in the spherical cap at the intersection of spherical surface over which satellites are distributed and the cone with side length $d_0$ centered at the location of the user/GW region closer to the user.
\par
Contact angle is defined as the minimum zenith Angle between the user and the satellite. As the distance between user and satellite increases, its zenith Angle also increases, and there is a unique correspondence between the two. Therefore, the contact Angle can be regarded as the representation of the contact distance in spherical coordinates. 
\par

\begin{figure}[h]
	\centering
	\includegraphics[width=0.95\linewidth]{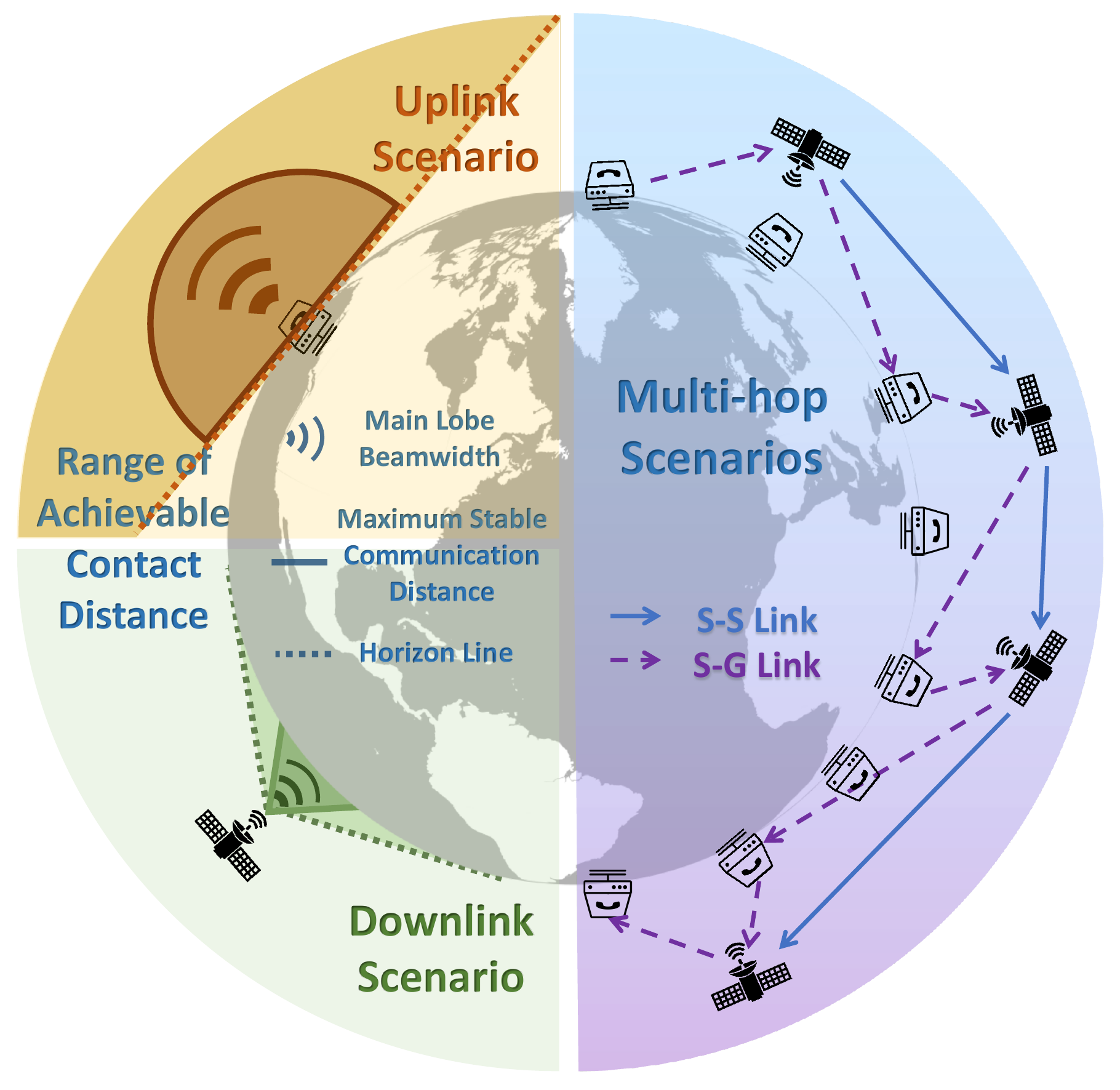}
	\caption{LEO satellite system structure diagram.}
	\label{fig:model}
\end{figure}
\par

From spatial distribution perspectives, satellites, GWs, and users are all located on spherical surfaces. We need to determine the range of achievable contact distance. The upper left region of Fig.~\ref{fig:model} shows the range of possible locations of a serving LEO satellite for a given location of the user/GW in the downlink. This range is mainly defined, as explained earlier, by the horizon, below which no LoS satellites are available \cite{Al_4}. When Beamforming (BF) is used, the main lobe beam-width is also one of the limitations \cite{OK_2}. The range of achievable contact distances is the intersection region above the horizon and in the beam's main-lobe width.
\par
Nearest neighbor is used to describe the shortest distance from a selected satellite to another. It is an important part of multi-hop network analysis. The difference between nearest neighbor and contact distance is that, the user/GW does not belong to the satellite point process, while the selected satellite in the nearest neighbor does \cite{MU_1}.
\par
When the satellites are deployed at different altitudes with different values of transmission power, the concept of \textit{association probability} needs to be introduced \cite{bushnaq2020optimal}. Especially in a multi-tier satellite network, it is reasonable to adopt different transmitting power for different altitude satellites. The satellite locations are divided into multiple point processes according to their altitudes and transmission power. For each point process, the distribution of the nearest satellite to the user needs to be obtained, which is exactly equivalent to obtaining the contact distance. The association probability with a given tier is the probability that the nearest satellite from this tier provides higher average received power than the that of the nearest satellites from all other tiers.
\par
Satellite availability means that at least one satellite is within the visible range of the user/GW. In small, low-altitude constellations, it is common to have no usable satellites above the horizon. In the existing literature, there are three methods to deal with the satellite availability problem: (\romannumeral1) reduce the probability of having no usable satellites below an acceptable threshold by increasing satellite density \cite{MU_1}, (\romannumeral2) enhance the probability of having available satellites by increasing the height and transmitting power simultaneously \cite{Al_4},\cite{OK_2}, and (\romannumeral3) calculate the probability of no available LoS satellite and set SINR to zero if no LoS satellites are available \cite{OK_1}. The first two proposals provide guidelines for constellation designers, and the last proposal provides the most accurate estimation of system performance for a designed constellation.

\begin{table*}[]
\caption{Models used in different literature.}
\renewcommand\arraystretch{0.2}
 \resizebox{18cm}{1.5 cm}{
\begin{tabular}{|c|c|c|c|c|c|}
\hline
{\color[HTML]{000000} References} &
  {\color[HTML]{000000} Scenarios} &
  {\color[HTML]{000000} Distribution of satellites} &
  {\color[HTML]{000000} Small scale fading} &
  {\color[HTML]{000000} System metrics} &
  {\color[HTML]{000000} System types} \\ \hline \hline 
{\color[HTML]{000000}  \cite{OK_1}, \cite{OK_3}, \cite{OK_2}} &
  {\color[HTML]{000000} Direct communication} &
  {\color[HTML]{000000} \begin{tabular}[c]{@{}c@{}} Non-homogeneous PPP\\    \\ BPP \end{tabular}}&
  {\color[HTML]{000000} \begin{tabular}[c]{@{}c@{}}Non-fading\\    \\ Rayleigh fading\\    \\ Rician fading\end{tabular}} &
  {\color[HTML]{000000} \begin{tabular}[c]{@{}c@{}}Coverage probability\\    \\ Average achievable rate\end{tabular}} &
  {\color[HTML]{000000} \begin{tabular}[c]{@{}c@{}}Noise-limited system\\    \\ Interference-limited system \\  \\  Generic system\end{tabular}} \\ \hline
{\color[HTML]{000000}  \cite{Al_1}, \cite{Al_4}} &
  {\color[HTML]{000000} Hybrid   network} &
  {\color[HTML]{000000} PPP} &
  {\color[HTML]{000000} Rayleigh fading} &
  {\color[HTML]{000000} Coverage probability} &
  {\color[HTML]{000000} \begin{tabular}[c]{@{}c@{}}Ideal system\\    \\ Noise-limited system\\    \\ Generic system\end{tabular}} \\ \hline
{\color[HTML]{000000}  \cite{MU_2}, \cite{MU_1}} &
  {\color[HTML]{000000} \begin{tabular}[c]{@{}c@{}}Multi-tier system\\    \\ GW-relayed\end{tabular}} &
  {\color[HTML]{000000} BPP} &
  {\color[HTML]{000000} Shadowed-Rician fading} &
  {\color[HTML]{000000} Coverage probability} &
  {\color[HTML]{000000} Noise-limited system} \\ \hline
\end{tabular}
    }
\label{table:models}
\end{table*}

\subsection{Coverage Probability and Average Achievable Rate}
In this section, we study the application of SG in system performance analysis, with emphasis on coverage probability. The definitions of the three metrics for measuring system performance are given below. 
\begin{itemize}
    \item The \textbf{\ac{SINR}} is used to measure the communication quality of the system. 
    \item \textbf{Coverage probability} is defined as the probability that the SINR is larger than a predetermined acceptable threshold. It represents the probability that the system can provide reliable connections.
    \item \textbf{Average achievable rate} is the ergodic capacity from the Shannon–Hartley theorem over a fading communication link.
\end{itemize}
\par
After the distance to the associated satellite is determined by the contact distance, the sum of the power of other LoS satellites is the interference.  SINR can be obtained then the coverage probability and average achievable rate are calculated. When LEO satellite adopts multiple orthogonal bands, the average achievable rate should be divided by the number of orthogonal bands.
\par
Coverage probabilities in different application scenarios are expressed differently. When LEO satellites rely on a ground GW as a relay, a user is covered by a satellite when both satellite-GW and GW-user links achieve the SINR requirements. Therefore, the system coverage probability should be the product of the coverage probability on both links. In the terrestrial base station and LEO satellite hybrid network scenario, satellites and BSs will independently establish the network. The user is considered in outage when the SINR of either the user-GW link or the GW-satellite link is below the threshold. 
\par
Coverage probabilities in different system types is  highlighted in the literature. In an ideal System, the signal will not be affected by any interference and noise in transmission. As long as there is a satellite in service, the user can always acquire a successful transmission. Such systems are used to study the upper bounds of performance and satellite availability. A noise-limited system occurs in an environment where the noise power is much stronger than the interference power. For example, when the frequency bands used by satellites are orthogonal or the interference comes from a diffraction signal, the interference power is often regarded as zero or a small constant. In addition, although using orthogonal channels will significantly reduce the Co-Channel Interference and improve the SINR and coverage probability, each satellite can use only a portion of the system's bandwidth, which reduces capacity and data rates. In an interference-limited system, interference power is dominant. Massive satellite constellations can cause huge interference. A generic system takes both interference and noise power into account.
\par
Coverage probabilities and average achievable rate under different types of Small scale fading have completely different mathematical expressions. Coverage probability and average achievable rate under Non-fading and Rayleigh in the generic system are provided in \cite{OK_1}. In Rician fading and SR fading, the associated satellite signal is transmitted in the direct path, while the interfering satellite signals are transmitted through the reflection path. In this case, the system is noise-limited. In \cite{MU_2}, the coverage probability under SR fading in a noise-limited system is obtained.
\par
Finally, no matter what kind of Small scale fading is adopted, the Laplace transform of interference is involved in the calculation of coverage probability and achievable rate \cite{OK_1}.

\section{SG-based Simulation Setup}
Suppose two GWs at opposite ends of the earth, with zenith angles $0$ and $\pi$ (not necessarily north and south poles), need to communicate. Satellites are modeled as a BPP on a sphere concentric with the earth. The signals of the inter-satellite link and the satellite-ground link travel at the speed of light. As is shown in Fig.~\ref{fig:model}, GWs can send messages to the associated satellites, while satellites can communicate with the GWs in the range of line-of-sight. We designed two scenarios \cite{chiariotti2020information}: (\romannumeral1) in \textit{no inter-satellite links scenario}, the satellite has to rely on the GW as a relay. In other words, the satellite uses GWs to forward data to other satellites until the message is delivered to its destination, (\romannumeral2) instead of requiring a GW relay, the satellite sends a message directly to the next hop in \textit{inter-satellite links scenario}. Examples of both scenarios are shown in Fig.~\ref{fig:model}.
\begin{figure}[h]
	\centering
	\includegraphics[width=0.95\linewidth]{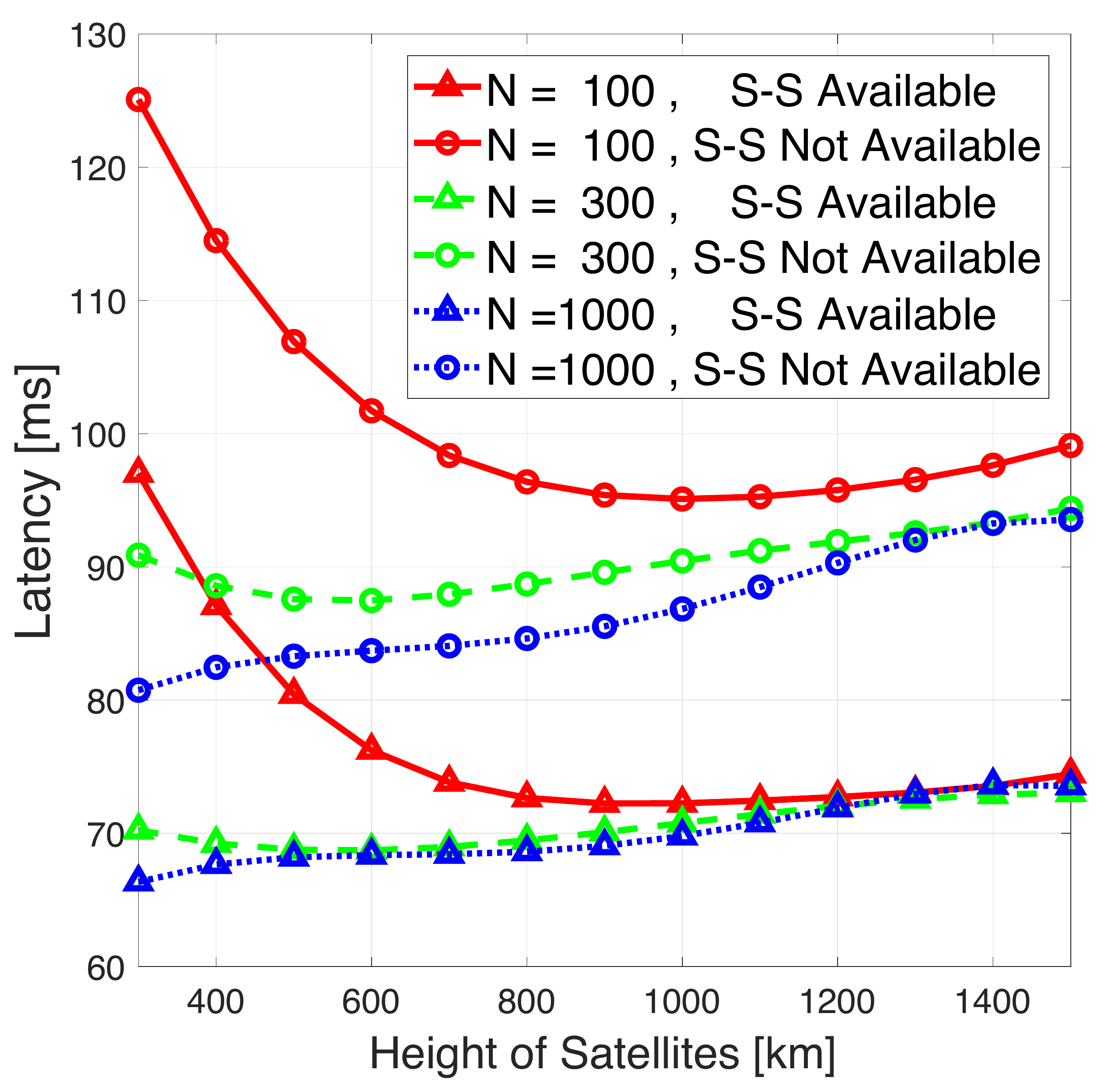}
	\caption{Average latency of satellite systems at different altitudes and scales.}
	\label{fig:latency}
\end{figure}
\par
In Fig.~\ref{fig:latency}, it can be seen that when the number of satellites is insufficient, the time delay decreases with the increase of altitude. That's because when a satellite is looking for the next hop, it can only try to find satellites in line of sight that might transmit signals. As it rises in altitude, the satellite's range of view expands, allowing it to choose which of several satellites is in a better position. But for enough satellites, with the increase of height, both the distances of inter-satellite link and satellite-ground link will increase. 
\par
Overall, when S-S communication is not available, the delay increases by about 30 percent compared to the case when S-S communication is available. When the number of satellites is increased from 100 to 300, the system delay can be significantly reduced. From 300 to 1,000 satellites, performance no longer improved much. Especially when S-S communication is not available, the curves of 300 satellites and 1000 satellites overlap, and adding additional satellites will not bring additional benefits in terms of delay. As the altitude increases, the two considered scenarios tend to have different latency values. The trends of the three curves when S-S communication is available tend to be consistent faster. It suggests that increasing altitude, increasing constellation size, and allowing interaction between satellites lead to more satellites within a satellite's reach. When there are enough satellites in the range, the performance improvement of these methods is limited, and the trend of the curves tends to be consistent.
\par
\begin{figure}[h]
	\centering
	\includegraphics[width=0.95\linewidth]{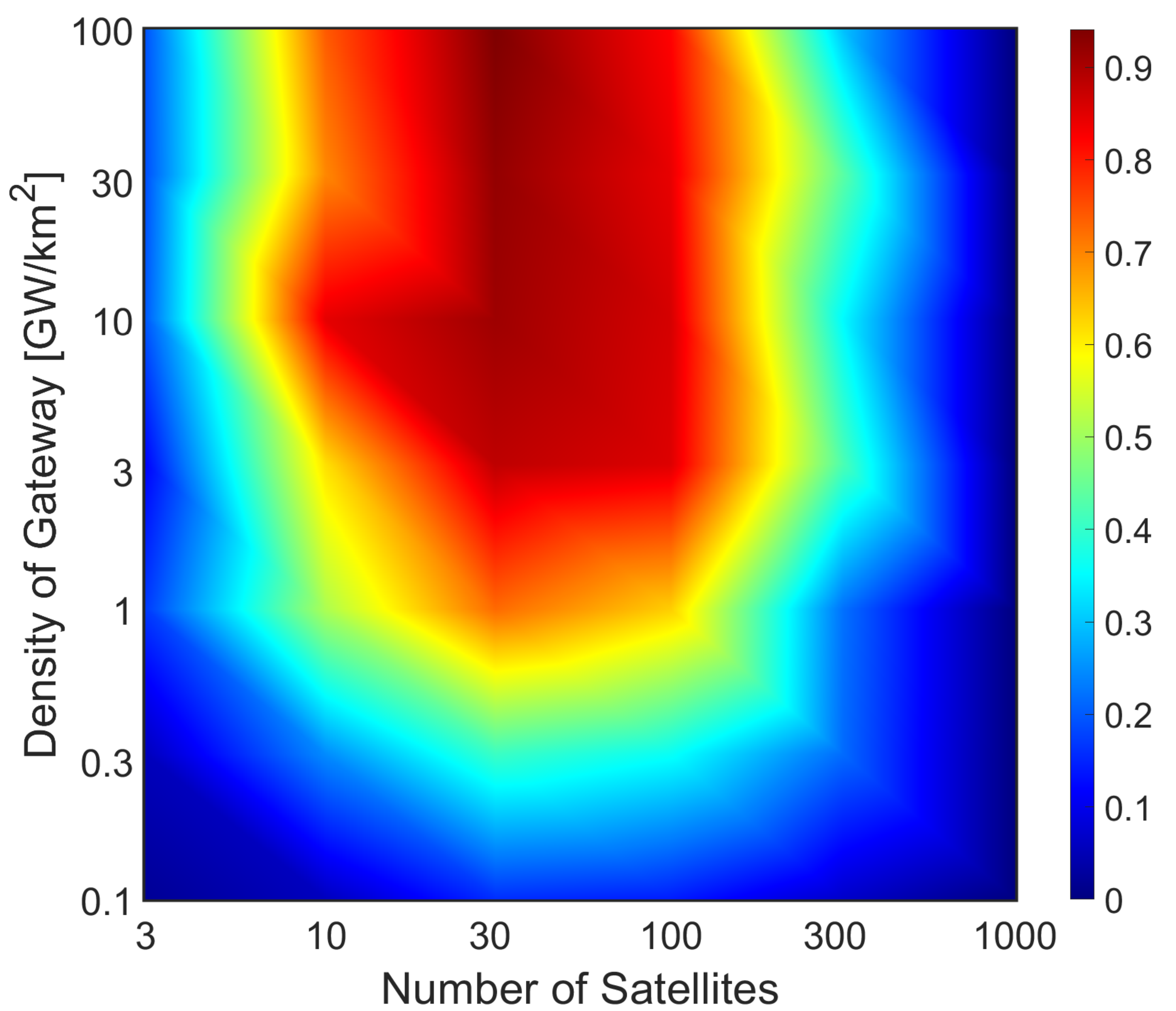}
	\caption{Influence of satellite number and GW density on coverage probability.}
	\label{fig:density}
\end{figure}
\begin{figure}[h]
	\centering
	\includegraphics[width=0.95\linewidth]{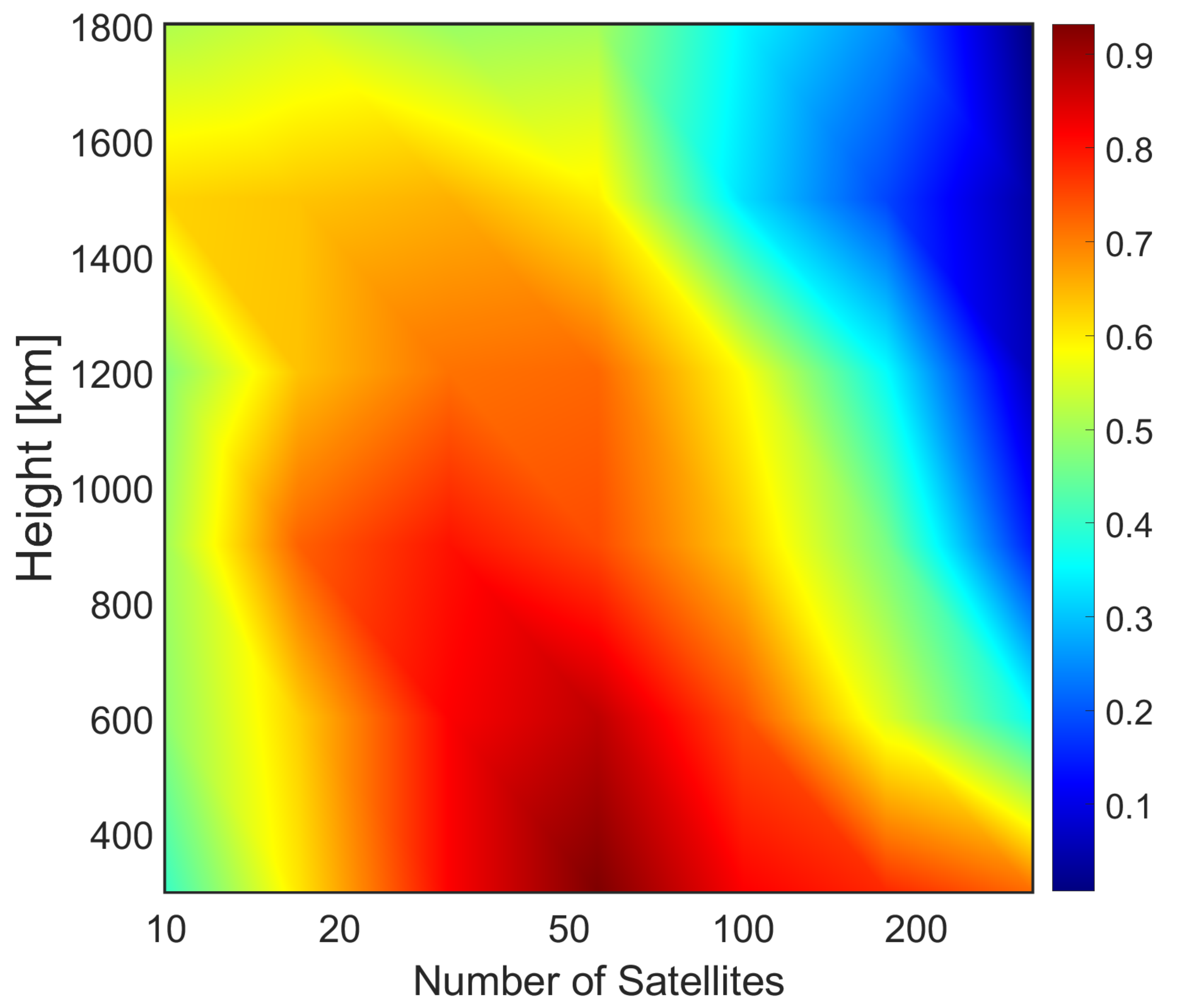}
	\caption{Influence of satellite number and constellation height on coverage probability.}
	\label{fig:height}
\end{figure}

In Fig.~\ref{fig:density} and Fig.~\ref{fig:height}, we analyze the influence of satellite number, GW density and constellation altitude on system coverage probability. Satellites follow BPP, and GWs follow PPP. We assume that the transmitting power of the satellite is 15dBw. The total coverage probability of the system is the product of the coverage probability of two links: (\romannumeral1) satellite-GW link and (\romannumeral2) GW-user link. The satellite-GW link and GW-user link follow the same channel fading, Large scale fading is 0dB in LoS range, Small scale fading follows $\mathcal{SR}(1.29,0.158,19.4)$ \cite{MU_2}, and the pass-loss exponent is 2. When calculating the coverage probability, the threshold of SINR is -10dB.
\par
In Fig.~\ref{fig:density}, the satellites have an altitude of 1000km, that is, they are distributed on a spherical surfaces with a radius of 7371km (radius of the earth is 6371km). When satellites are insufficient, the system lacks available satellites to provide coverage, or the associated satellites are too far away. Conversely, too many satellites can cause significant interference. Therefore, with the increase of the number of satellites, the coverage probability shows a trend of first increasing and then decreasing. In addition, at an altitude of 1000km, the optimal number of satellites for coverage probability is about 30. The optimal number of satellites is independent of the density of GWs. Unlike satellites, increasing the density of GWs in a reasonable range will improve the coverage performance of the system, and the GW subsystem shows the characteristics of a noise-limited system.
y

\par
In Fig.~\ref{fig:height}, the density of the GW is fixed at 3/km$^2$. The increase of constellation height will reduce the coverage probability. There exists an optimal number of satellites corresponding to altitude. With the increase of constellation height, the number of satellites corresponding to the optimal coverage decreases.

\section{Challenges and Future scenarios}
Although SG has many advantages in system performance analysis, the traditional SG analysis framework has some limitations, and there are many challenges in applying SG to more complex and sophisticated satellite systems in the future. Therefore, in the right part of Fig.~\ref{fig:roadmap}, we envision several future scenarios. These scenarios can also be modeled, analyzed, and evaluated using SG-based models as described in this article.
\par
\textit{S-S Communication}: In the previous analysis, we assumed that satellites were independent of each other and analyzed the lower bound of the system performance. But in future satellite systems, there will be more S-S Communications, that will significantly improve the system's performance. Through interaction, satellites can learn more about the environment, such as the number of nearby satellites and the frequency bands they use. communicating with satellites in front of a predetermined trajectory allows them to predict future communication environments. The problem of insufficient capacity and achievable data rate can be effectively alleviated by smartly designing association and hand over techniques between satellites. Conversely, with the application of orthogonal channels, it is possible to have a high SINR even if the user/GW is associated with a non-optimal satellite after the handover.
\par
\textit{Self-adapting to the environment}: The second challenge faced by traditional SG analysis frameworks is that, in reality, satellites do not remain inactive when communications fail in harsh environments. While facing different astronomical and geographical conditions, actively perceiving the environment and selecting proper adaptive strategies are important. For example, the satellite can always point the main lobe to the area with the greatest demand for services, greatly improving local coverage. In the face of heavy rain attenuation, satellites are expected to increase the transmission power to maintain an acceptable SINR. 
\par
\textit{Virtual Satellite}: In addition to S-S communications, there may be direct cooperation between satellites. Similar to the idea of virtual hosting, Multiple satellites with the same mission can be integrated into a virtual satellite. The virtual satellite can decide the number of orthogonal channels and allocate specific frequencies internally. What's more, LEO satellites can be used as the auxiliary of GEO satellites to complement the coverage of regions that a single GEO cannot meet independently.

\section{Conclusions}
This paper investigates the application of SG to simulating and analyzing LEO satellite communications systems, introduces several satellite distributions and analyzes the similarities and differences of channel models in the literature. From a non-technical perspective, we describe coverage probability in different scenarios, average achievable rate, and contact distance. This paper analyzes several factors affecting latency and coverage probability using SG-based models and gives the following simple conclusions. Increasing the number of satellites and the height of the constellation within a specific range can effectively reduce the latency. The ability of the satellites to communicate with each other has a huge effect on reducing the value of latency by 30 percent. This indicates that the recent technological advances in enabling S-S communications will significantly widen the set of applications that can rely on LEO satellite communications. For LEO satellites, there is an optimal number of satellites to achieve the best coverage performance. The optimal number of satellites decreases with constellation height and is independent of the GW density. Increasing the density of the GW can effectively improve the system coverage while increasing the height will reduce the system coverage when the number of satellites remains unchanged.
\input{references.bbl}

\bibliographystyle{IEEEtran}
\bibliography{references}

\section*{biographies}
\begin{IEEEbiographynophoto}
{Ruibo Wang} is a Master candidate at King Abdullah University of Science and Technology (KAUST). He received his B.S. degree in Communication Engineering from University of Electronic Science and Technology of China in 2020.
\end{IEEEbiographynophoto}
\begin{IEEEbiographynophoto}
{Mustafa A. Kishk} [S’16, M’18] is a postdoctoral research fellow at KAUST. He received his Ph.D. degree in electrical engineering from Virginia Tech in 2018.
\end{IEEEbiographynophoto}
\begin{IEEEbiographynophoto}
{Mohamed-Slim Alouini} [S’94, M’98, SM’03, F’09] is a professor at KAUST. He received his Ph.D. degree in electrical engineering from California Institute of Technology in 1998.
\end{IEEEbiographynophoto}
\end{document}

%% file: notation.tex
\def\nb0{{\mathbf{0}}}
\def\nb1{{\mathbf{1}}}



%% file: Elzanaty_Acronyms.tex
\begin{acronym}

\acro{5G-NR}{5G New Radio}
\acro{3GPP}{3rd Generation Partnership Project}
\acro{ABS}{aerial base station}
\acro{AC}{address coding}
\acro{ACF}{autocorrelation function}
\acro{ACR}{autocorrelation receiver}
\acro{ADC}{analog-to-digital converter}
\acrodef{aic}[AIC]{Analog-to-Information Converter}     
\acro{AIC}[AIC]{Akaike information criterion}
\acro{aric}[ARIC]{asymmetric restricted isometry constant}
\acro{arip}[ARIP]{asymmetric restricted isometry property}

\acro{ARQ}{Automatic Repeat Request}
\acro{AUB}{asymptotic union bound}
\acrodef{awgn}[AWGN]{Additive White Gaussian Noise}     
\acro{AWGN}{additive white Gaussian noise}

\acro{APSK}[PSK]{asymmetric PSK} 

\acro{waric}[AWRICs]{asymmetric weak restricted isometry constants}
\acro{warip}[AWRIP]{asymmetric weak restricted isometry property}
\acro{BCH}{Bose, Chaudhuri, and Hocquenghem}        
\acro{BCHC}[BCHSC]{BCH based source coding}
\acro{BEP}{bit error probability}
\acro{BFC}{block fading channel}
\acro{BG}[BG]{Bernoulli-Gaussian}
\acro{BGG}{Bernoulli-Generalized Gaussian}
\acro{BPAM}{binary pulse amplitude modulation}
\acro{BPDN}{Basis Pursuit Denoising}
\acro{BPPM}{binary pulse position modulation}
\acro{BPSK}{Binary Phase Shift Keying}
\acro{BPZF}{bandpass zonal filter}
\acro{BSC}{binary symmetric channels}              
\acro{BU}[BU]{Bernoulli-uniform}
\acro{BER}{bit error rate}
\acro{BS}{base station}
\acro{BW}{BandWidth}
\acro{BLLL}{ binary log-linear learning }

\acro{CP}{Cyclic Prefix}
\acrodef{cdf}[CDF]{cumulative distribution function}   
\acro{CDF}{Cumulative Distribution Function}
\acrodef{c.d.f.}[CDF]{cumulative distribution function}
\acro{CCDF}{complementary cumulative distribution function}
\acrodef{ccdf}[CCDF]{complementary CDF}               
\acrodef{c.c.d.f.}[CCDF]{complementary cumulative distribution function}
\acro{CD}{cooperative diversity}

\acro{CDMA}{Code Division Multiple Access}
\acro{ch.f.}{characteristic function}
\acro{CIR}{channel impulse response}
\acro{cosamp}[CoSaMP]{compressive sampling matching pursuit}
\acro{CR}{cognitive radio}
\acro{cs}[CS]{compressed sensing}                   
\acrodef{cscapital}[CS]{Compressed sensing} 
\acrodef{CS}[CS]{compressed sensing}
\acro{CSI}{channel state information}
\acro{CCSDS}{consultative committee for space data systems}
\acro{CC}{convolutional coding}
\acro{Covid19}[COVID-19]{Coronavirus disease}

\acro{DAA}{detect and avoid}
\acro{DAB}{digital audio broadcasting}
\acro{DCT}{discrete cosine transform}
\acro{dft}[DFT]{discrete Fourier transform}
\acro{DR}{distortion-rate}
\acro{DS}{direct sequence}
\acro{DS-SS}{direct-sequence spread-spectrum}
\acro{DTR}{differential transmitted-reference}
\acro{DVB-H}{digital video broadcasting\,--\,handheld}
\acro{DVB-T}{digital video broadcasting\,--\,terrestrial}
\acro{DL}{DownLink}
\acro{DSSS}{Direct Sequence Spread Spectrum}
\acro{DFT-s-OFDM}{Discrete Fourier Transform-spread-Orthogonal Frequency Division Multiplexing}
\acro{DAS}{Distributed Antenna System}
\acro{DNA}{DeoxyriboNucleic Acid}

\acro{EC}{European Commission}
\acro{EED}[EED]{exact eigenvalues distribution}
\acro{EIRP}{Equivalent Isotropically Radiated Power}
\acro{ELP}{equivalent low-pass}
\acro{eMBB}{Enhanced Mobile Broadband}
\acro{EMF}{ElectroMagnetic Field}
\acro{EU}{European union}
\acro{EI}{Exposure Index}
\acro{eICIC}{enhanced Inter-Cell Interference Coordination}

\acro{FC}[FC]{fusion center}
\acro{FCC}{Federal Communications Commission}
\acro{FEC}{forward error correction}
\acro{FFT}{fast Fourier transform}
\acro{FH}{frequency-hopping}
\acro{FH-SS}{frequency-hopping spread-spectrum}
\acrodef{FS}{Frame synchronization}
\acro{FSsmall}[FS]{frame synchronization}  
\acro{FDMA}{Frequency Division Multiple Access}

\acro{GA}{Gaussian approximation}
\acro{GF}{Galois field }
\acro{GG}{Generalized-Gaussian}
\acro{GIC}[GIC]{generalized information criterion}
\acro{GLRT}{generalized likelihood ratio test}
\acro{GPS}{Global Positioning System}
\acro{GMSK}{Gaussian Minimum Shift Keying}
\acro{GSMA}{Global System for Mobile communications Association}
\acro{GS}{ground station}
\acro{GMG}{ Grid-connected MicroGeneration}

\acro{HAP}{high altitude platform}
\acro{HetNet}{Heterogeneous network}

\acro{IDR}{information distortion-rate}
\acro{IFFT}{inverse fast Fourier transform}
\acro{iht}[IHT]{iterative hard thresholding}
\acro{i.i.d.}{independent, identically distributed}
\acro{IoT}{Internet of Things}                      
\acro{IR}{impulse radio}
\acro{lric}[LRIC]{lower restricted isometry constant}
\acro{lrict}[LRICt]{lower restricted isometry constant threshold}
\acro{ISI}{intersymbol interference}
\acro{ITU}{International Telecommunication Union}
\acro{ICNIRP}{International Commission on Non-Ionizing Radiation Protection}
\acro{IEEE}{Institute of Electrical and Electronics Engineers}
\acro{ICES}{IEEE international committee on electromagnetic safety}
\acro{IEC}{International Electrotechnical Commission}
\acro{IARC}{International Agency on Research on Cancer}
\acro{IS-95}{Interim Standard 95}

\acro{KPI}{Key Performance Indicator}

\acro{LEO}{low earth orbit}
\acro{LF}{likelihood function}
\acro{LLF}{log-likelihood function}
\acro{LLR}{log-likelihood ratio}
\acro{LLRT}{log-likelihood ratio test}
\acro{LoS}{Line-of-Sight}
\acro{LRT}{likelihood ratio test}
\acro{wlric}[LWRIC]{lower weak restricted isometry constant}
\acro{wlrict}[LWRICt]{LWRIC threshold}
\acro{LPWAN}{Low Power Wide Area Network}
\acro{LoRaWAN}{Low power long Range Wide Area Network}
\acro{NLoS}{Non-Line-of-Sight}
\acro{LiFi}[Li-Fi]{light-fidelity}
 \acro{LED}{light emitting diode}
 \acro{LABS}{LoS transmission with each ABS}
 \acro{NLABS}{NLoS transmission with each ABS}

\acro{MB}{multiband}
\acro{MC}{macro cell}
\acro{MDS}{mixed distributed source}
\acro{MF}{matched filter}
\acro{m.g.f.}{moment generating function}
\acro{MI}{mutual information}
\acro{MIMO}{Multiple-Input Multiple-Output}
\acro{MISO}{multiple-input single-output}
\acrodef{maxs}[MJSO]{maximum joint support cardinality}                       
\acro{ML}[ML]{maximum likelihood}
\acro{MMSE}{minimum mean-square error}
\acro{MMV}{multiple measurement vectors}
\acrodef{MOS}{model order selection}
\acro{M-PSK}[${M}$-PSK]{$M$-ary phase shift keying}                       
\acro{M-APSK}[${M}$-PSK]{$M$-ary asymmetric PSK} 
\acro{MP}{ multi-period}
\acro{MINLP}{mixed integer non-linear programming}

\acro{M-QAM}[$M$-QAM]{$M$-ary quadrature amplitude modulation}
\acro{MRC}{maximal ratio combiner}                  
\acro{maxs}[MSO]{maximum sparsity order}                                      
\acro{M2M}{Machine-to-Machine}                                                
\acro{MUI}{multi-user interference}
\acro{mMTC}{massive Machine Type Communications}      
\acro{mm-Wave}{millimeter-wave}
\acro{MP}{mobile phone}
\acro{MPE}{maximum permissible exposure}
\acro{MAC}{media access control}
\acro{NB}{narrowband}
\acro{NBI}{narrowband interference}
\acro{NLA}{nonlinear sparse approximation}
\acro{NLOS}{Non-Line of Sight}
\acro{NTIA}{National Telecommunications and Information Administration}
\acro{NTP}{National Toxicology Program}
\acro{NHS}{National Health Service}

\acro{LOS}{Line of Sight}

\acro{OC}{optimum combining}                             
\acro{OC}{optimum combining}
\acro{ODE}{operational distortion-energy}
\acro{ODR}{operational distortion-rate}
\acro{OFDM}{Orthogonal Frequency-Division Multiplexing}
\acro{omp}[OMP]{orthogonal matching pursuit}
\acro{OSMP}[OSMP]{orthogonal subspace matching pursuit}
\acro{OQAM}{offset quadrature amplitude modulation}
\acro{OQPSK}{offset QPSK}
\acro{OFDMA}{Orthogonal Frequency-division Multiple Access}
\acro{OPEX}{Operating Expenditures}
\acro{OQPSK/PM}{OQPSK with phase modulation}

\acro{PAM}{pulse amplitude modulation}
\acro{PAR}{peak-to-average ratio}
\acrodef{pdf}[PDF]{probability density function}                      
\acro{PDF}{probability density function}
\acrodef{p.d.f.}[PDF]{probability distribution function}
\acro{PDP}{power dispersion profile}
\acro{PMF}{probability mass function}                             
\acrodef{p.m.f.}[PMF]{probability mass function}
\acro{PN}{pseudo-noise}
\acro{PPM}{pulse position modulation}
\acro{PRake}{Partial Rake}
\acro{PSD}{power spectral density}
\acro{PSEP}{pairwise synchronization error probability}
\acro{PSK}{phase shift keying}
\acro{PD}{power density}
\acro{8-PSK}[$8$-PSK]{$8$-phase shift keying}
\acro{PPP}{Poisson point process}
\acro{PCP}{Poisson cluster process}
 
\acro{FSK}{Frequency Shift Keying}

\acro{QAM}{Quadrature Amplitude Modulation}
\acro{QPSK}{Quadrature Phase Shift Keying}
\acro{OQPSK/PM}{OQPSK with phase modulator }

\acro{RD}[RD]{raw data}
\acro{RDL}{"random data limit"}
\acro{ric}[RIC]{restricted isometry constant}
\acro{rict}[RICt]{restricted isometry constant threshold}
\acro{rip}[RIP]{restricted isometry property}
\acro{ROC}{receiver operating characteristic}
\acro{rq}[RQ]{Raleigh quotient}
\acro{RS}[RS]{Reed-Solomon}
\acro{RSC}[RSSC]{RS based source coding}
\acro{r.v.}{random variable}                               
\acro{R.V.}{random vector}
\acro{RMS}{root mean square}
\acro{RFR}{radiofrequency radiation}
\acro{RIS}{Reconfigurable Intelligent Surface}
\acro{RNA}{RiboNucleic Acid}
\acro{RRM}{Radio Resource Management}
\acro{RUE}{reference user equipments}
\acro{RAT}{radio access technology}
\acro{RB}{resource block}

\acro{SA}[SA-Music]{subspace-augmented MUSIC with OSMP}
\acro{SC}{small cell}
\acro{SCBSES}[SCBSES]{Source Compression Based Syndrome Encoding Scheme}
\acro{SCM}{sample covariance matrix}
\acro{SEP}{symbol error probability}
\acro{SG}[SG]{sparse-land Gaussian model}
\acro{SIMO}{single-input multiple-output}
\acro{SINR}{signal-to-interference plus noise ratio}
\acro{SIR}{signal-to-interference ratio}
\acro{SISO}{Single-Input Single-Output}
\acro{SMV}{single measurement vector}
\acro{SNR}[\textrm{SNR}]{signal-to-noise ratio} 
\acro{sp}[SP]{subspace pursuit}
\acro{SS}{spread spectrum}
\acro{SW}{sync word}
\acro{SAR}{specific absorption rate}
\acro{SSB}{synchronization signal block}
\acro{SR}{shrink and realign}

\acro{tUAV}{tethered Unmanned Aerial Vehicle}
\acro{TBS}{terrestrial base station}

\acro{uUAV}{untethered Unmanned Aerial Vehicle}
\acro{PDF}{probability density functions}

\acro{PL}{path-loss}

\acro{TH}{time-hopping}
\acro{ToA}{time-of-arrival}
\acro{TR}{transmitted-reference}
\acro{TW}{Tracy-Widom}
\acro{TWDT}{TW Distribution Tail}
\acro{TCM}{trellis coded modulation}
\acro{TDD}{Time-Division Duplexing}
\acro{TDMA}{Time Division Multiple Access}
\acro{Tx}{average transmit}

\acro{UAV}{Unmanned Aerial Vehicle}
\acro{uric}[URIC]{upper restricted isometry constant}
\acro{urict}[URICt]{upper restricted isometry constant threshold}
\acro{UWB}{ultrawide band}
\acro{UWBcap}[UWB]{Ultrawide band}   
\acro{URLLC}{Ultra Reliable Low Latency Communications}
         
\acro{wuric}[UWRIC]{upper weak restricted isometry constant}
\acro{wurict}[UWRICt]{UWRIC threshold}                
\acro{UE}{User Equipment}
\acro{UL}{UpLink}

\acro{WiM}[WiM]{weigh-in-motion}
\acro{WLAN}{wireless local area network}
\acro{wm}[WM]{Wishart matrix}                               
\acroplural{wm}[WM]{Wishart matrices}
\acro{WMAN}{wireless metropolitan area network}
\acro{WPAN}{wireless personal area network}
\acro{wric}[WRIC]{weak restricted isometry constant}
\acro{wrict}[WRICt]{weak restricted isometry constant thresholds}
\acro{wrip}[WRIP]{weak restricted isometry property}
\acro{WSN}{wireless sensor network}                        
\acro{WSS}{Wide-Sense Stationary}
\acro{WHO}{World Health Organization}
\acro{Wi-Fi}{Wireless Fidelity}

\acro{sss}[SpaSoSEnc]{sparse source syndrome encoding}

\acro{VLC}{Visible Light Communication}
\acro{VPN}{Virtual Private Network} 
\acro{RF}{Radio Frequency}
\acro{FSO}{Free Space Optics}
\acro{IoST}{Internet of Space Things}

\acro{GSM}{Global System for Mobile Communications}
\acro{2G}{Second-generation cellular network}
\acro{3G}{Third-generation cellular network}
\acro{4G}{Fourth-generation cellular network}
\acro{5G}{Fifth-generation cellular network}	
\acro{gNB}{next-generation Node-B Base Station}
\acro{NR}{New Radio}
\acro{UMTS}{Universal Mobile Telecommunications Service}
\acro{LTE}{Long Term Evolution}

\acro{QoS}{Quality of Service}
\end{acronym}

%% file: Definitions.tex
\newcommand{\SAR} {\mathrm{SAR}}
\newcommand{\WBSAR} {\mathrm{SAR}_{\mathsf{WB}}}
\newcommand{\gSAR} {\mathrm{SAR}_{10\si{\gram}}}
\newcommand{\Sab} {S_{\mathsf{ab}}}
\newcommand{\Eavg} {E_{\mathsf{avg}}}
\newcommand{\ft}{f_{\textsf{th}}}
\newcommand{\alphatf}{\alpha_{24}}